# Drastic slowdown of shear waves in unjammed granular suspensions

J. Brum[*], J.-L. Gennisson[†], M. Fink, A. Tourin, and X. Jia[§]

Institut Langevin, ESPCI Paris, PSL University, CNRS, 1 rue Jussieu, 75005 Paris, France

**Abstract**

We present an experimental investigation of shear elastic wave propagation along the surface of a dense granular suspension. Using an ultrafast ultrasound scanner, we monitor the softening of the shear wave velocity inside the optically opaque medium as the driving amplitude increases. For such nonlinear behavior two regimes are found: in the first regime, we observe a significant shear modulus weakening, but without visible grain rearrangements. In the second regime, there is a clear grain rearrangement accompanied by a modulus decrease up to 88%. A friction model is proposed to describe the interplay between nonlinear elasticity and plasticity, which highlights the crucial effect of contact slipping before contact breaking. Investigation of these nonlinear shear waves may bridge the gap between two disjoint approaches for describing the dynamics near unjamming: linear elastic soft modes and nonlinear collisional shock.

**Introduction**

The jamming transition is a general paradigm for understanding how complex fluids such as foams, emulsions, and granular materials develop rigidity: when the density of randomly packed particles is increased to a certain critical value, the viscosity increases dramatically and the flow arrests[1-6]. Reciprocally, amorphous solids made of athermal particles like bubbles, droplets, and grains lose shear rigidity and make a transition to a liquid state when the confining pressure vanishes[7]. Numerical simulations of frictionless particles show that an effective medium description fails near unjamming due to non-affine motion of particles and that the critical scaling of the shear modulus is correlated to soft modes[8-10]. However, understanding the mechanical response across this solid-like-to-liquid-like transition still remains a major challenge for real granular matter because of friction and because of the strong nonlinearity at vanishing confining pressure[11-13] where the particle packing does not tend to isostaticity[4,10]. Addressing this issue is also of great importance for industrial applications and geophysical processes such as landslides[14].

---

[*] Currently at Instituto de Fisica, Universidad de la República, Montevideo, Uruguay
[†] Currently at IR4M, CNRS, University Paris Saclay, CEA SHFJ, Orsay, France
[§] Corresponding author: xiaoping.jia@espci.fr

In both dry and water-saturated granular materials, force transmission and elastic wave propagation strongly depend on the inhomogeneous and metastable contact force network[2,5,8]. As the confining pressure $P$ decreases, the effective medium theory (EMT) [8] based on the affine approximation and the Hertz-Mindlin contact law predicts that in the linear regime both bulk $K$ and shear modulus $G$ scale as $\sim P^{1/3}$. However, numerical simulations in frictionless sphere packings reveal anomalous scaling of the shear modulus $G$ with vanishing pressure as $\sim P^{2/3}$ due to the nonaffine deformation[4,7-9]. Nonlinear responses beyond linear elasticity have also been investigated by using the finite shear strain to study the transition from a jammed to a flowing state[15-17].

Sound propagation in granular media provides a very efficient and controlled way to perform dynamic measurements that can be compared naturally with theory and simulations based on elasticity. The long-wavelength coherent wave gives access to the effective modulus whereas the short-wavelength scattered waves are sensitive to any rearrangement of the contact force network[19]. In the linear regime, velocity measurements of coherent sound waves allow monitoring of the weakening of jammed media when the confining pressure is decreased[8,19,20] or when a static shear is applied[20,21]. In the nonlinear regime, the high-amplitude ultrasound can act as a pump to soften the jammed solid as shown by using compressional wave[22-24]. Another dynamic approach using shock wave has also been proposed to investigate the unjamming transition in granular media[15]. In such highly nonlinear regime, the dynamic displacement is larger than the grain overlap induced by the confining pressure so that the elastic wave propagation becomes impossible, i.e. in sonic vacuum[16]. Instead, soliton-like shocks travel via collisions with a front speed depending on the particle velocity.

In this work, we investigate the unjamming transition by new measurements of high-amplitude *shear wave* in a weakly jammed granular suspension. Unlike oscillatory rheological measurements[17], these acoustic measurements allow us to monitor *locally* the shear modulus softening as the driving amplitude increases, along the wave path and inside optically opaque dense granular suspensions, till the onset of the plastic rearrangement of grains. We analyze these nonlinear elastic responses using mean-field descriptions for frictional sphere packings. Shear elastic waves are used here both as a pump to fluidize the granular solid and a probe of the material softening.

**Experiment**

Weakly jammed granular media under investigation are made with glass beads (diameter $d \sim$ 250 µm) confined in a rectangular box with a free surface (Fig. 1a). The glass beads settle down in water under gravity, which creates a dense granular suspension ($h \approx 13$ cm in height), with a packing density of about 60%. A rough metallic plate ($\Sigma_0 \approx 10 \times 10$ cm$^2$) glued with sand particles is used as a quasi-plane shear wave source. It is excited by a shaker with a four-cycle tone-burst centered at a frequency ranging from 100 to 500 Hz. The static load applied to the plate, $W = P\Sigma_0 \approx 7.5$ N, can be estimated from the mean confining pressure $P = (\rho_g - \rho_w)gL/2 \approx 750$ Pa with $\rho_g = 2500$ kg/m$^3$ and $\rho_w = 1000$ kg/m$^3$ the density of glass and water respectively, and $L \approx 10$ cm. Oscillating shear force $F_{ac}$ and acceleration $a_{ac}$ are measured by a force sensor and an accelerometer, respectively.

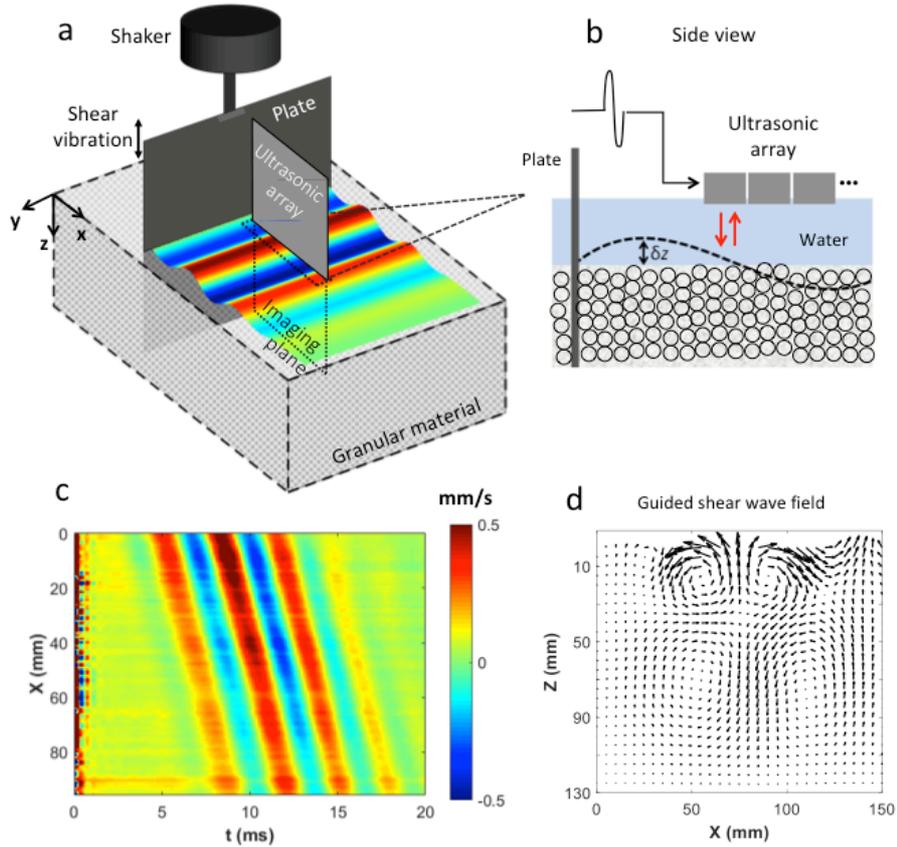

Fig. 1 Experimental investigation of a Rayleigh-like shear wave along the surface of a granular medium. (a) Sketch of the experimental setup: the wave is excited by a rough plate in a water-saturated glass bead packing. (b) The out-of-plane particle displacement ($\delta z$) and velocity at the sample surface is inferred from the cross correlation of successive backscattered ultrasonic speckles acquired with an ultrafast ultrasonic scanner. (c) Typical seismogram measured after the generation of a shear pulse centred at 300 Hz; the axial particle velocity is plotted versus time $t$ and distance $x$ from the source. (d) Snapshot of the vector displacement field of the Rayleigh-like surface mode calculated with free surface and clamped bottom boundary conditions in an inhomogeneous layer (see the Supplemental Material for details).

As shown in Fig. 1d, the source excites a Rayleigh-like surface wave[11, 27] with a group velocity close to the shear wave velocity $V_S$ (see the Supplemental Material for details). To investigate the propagation of this low-frequency shear guided mode we used an ultrafast ultrasound scanner (Aixplorer®) that was originally developed in our laboratory to track tissue motion induced by low-speed shear waves in the context of medical imaging[28]. This kind of scanner can acquire images up to 200 times faster than conventional ultrasound systems and it was also applied for rheology measurements in complex fluids[29].

Fig. 1b depicts a 192-element ultrasonic array (centred at 4 MHz) placed in water close to the surface of the granular suspension and parallel to the propagation direction ($x$-axis). After the generation of the guided wave, the ultrafast ultrasonic scanner acquires successive backscattered ultrasonic speckle patterns from the granular sample with a frame rate of $f_{RF}$ =16 kHz. The arrival time of a given speckle pattern corresponds to a specific location of grains within the medium. By cross-correlating in time and space the speckle observed from one frame to the next (i.e., speckle interferometry[29]), a speckle-tracking algorithm allows for the estimation of particle velocities along the ultrasonic beam direction $v_z = \delta z/\delta t$ with $\delta t = 1/f_{RF}$ (see Fig. 1b and the Supplemental Material). Fig. 1c shows a typical resulting seismogram at the surface of the bead packing ($z_0 = 0$) after a source excitation at a central frequency of $f = 300$ Hz and with small amplitude $F_{ac} \sim 0.19$ N. The measured particle velocity along z-axis is of the order of $v_{ac} \sim 0.4$ mm/s, which corresponds to a particle displacement $u_{ac}$ (= $v_{ac}/(2\pi f)$) ~ 0.3 µm, i.e., much less than the particle diameter. The shear pulse group velocity, $V_G \approx 25$ m/s, is determined from the slope of the seismogram. This gives a dynamic strain $\gamma_{ac}$ (= $\partial u_{ac}/\partial x = (2\pi f/V)u_{ac}$) ~ $2.10^{-5}$ with $V \sim V_G$ the phase velocity.

The main goal of our work was to monitor the shear velocity softening near unjamming in a realistic granular medium (i.e., frictional and optically opaque). When the driving amplitude is increased, a significant increase of the shear pulse time-of-flight is clearly observed (Fig. 2a), corresponding to a softening of the group wave velocity $V_G$ up to 30-50% as shown in Fig. 2b. Such a velocity softening is about 3 times larger than that observed with the compressional wave[25]. We also detect the generation of $2^{nd}$ and $3^{rd}$ harmonics whose amplitudes evolve as a function of the propagation distance (Fig. 2c). This observation is different from the ones with shear shock waves propagating in soft solids (like gels) where only odd harmonics are observed but without any significant change of the wave velocity (< 1%) [28].

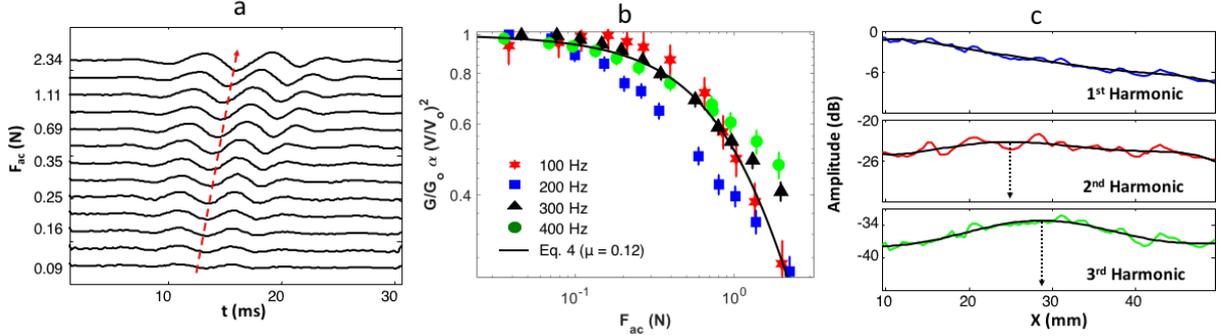

Fig. 2 Nonlinear acoustic responses of a shear wave in a fragile granular packing. (a) Shear wave pulse recorded at $x = 95.5$ mm from the source; the travel time of the shear pulse increases with the shear driving force $F_{ac}$. (b) Softening of the shear modulus $G$ (or group velocity $V_G$) as a function of the driving force $F_{ac}$ in a case when no visible motion of grains is observed. Solid line is the prediction by the friction model. (c) Amplitudes of the fundamental (100 Hz) and of the 2nd and 3rd harmonics versus propagation distance after an excitation at $F_{ac} \sim 2N$. The maximum is indicated by an arrow after smoothing (solid lines).

Notice that despite the important softening of the wave velocity $\Delta V_G/V_G$ (up to 30-50%) (Fig. 2b) presumably due to the modification of the contact network[23], we do not observe any visible rearrangement of grain positions. A similar behavior was observed in another kind of dynamic experiment where a dense granular suspension was subject to a sinusoidal oscillation with comparable shaking amplitude and frequency[30]. The absence of plastic rearrangement of grains can be explained as follows: on the one hand, the typical acoustic displacement is relatively small compared to one grain diameter i.e., $u_{ac} < 5$ μm ~ $d/50$). On the other hand, the characteristic time for rearrangement (measured by the time it takes for a grain to move from one cage to the next one over a distance ~ $d$) is large compared to the period of driving. More precisely, we find that the fall time for the granular suspension is $t_{fall}$ [$= \eta_f /(P_g \alpha)$] ~ 20 ms at a confining pressure $P_g \sim d \rho_p g \sim 2.5$ Pa (the layer at the top surface is the most likely to be subjected to possible grain motion[31]). For the vibration frequency range explored in our experiments ($f$ = 100-500 Hz), the period of oscillatory driving is $T_0 = 1/f \sim 4$ ms, which leads to $T_0$ smaller than $t_{fall}$ *(saturated)*. Under such conditions, we expect that the grains in the granular suspension do not have the time to move or rearrange to the new cages before the applied vibration changes the driving direction.

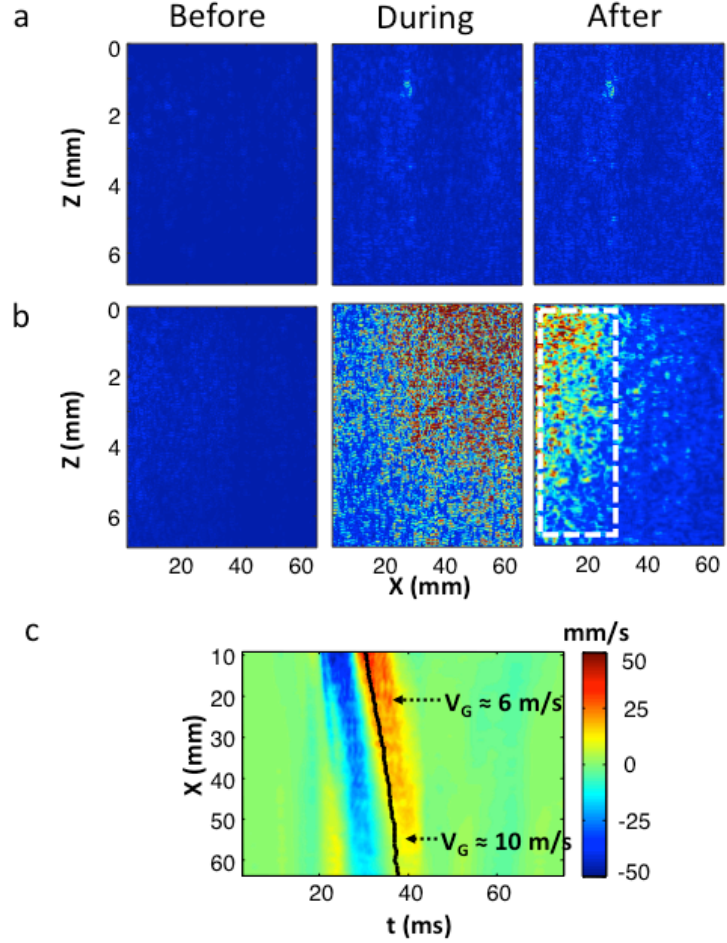

Fig. 3 Ultrasound imaging of grain motion during plastic granular rearrangement. Comparison of the ultrasonic speckle patterns recorded before, during and after the shear wave passage after an excitation at 100 Hz by (a) a small force $F_{ac} \approx 0.03$ N and (b) a large one $F_{ac} \approx 2.7$ N. This latter induces the rearrangement of grains localized in a zone close to the driving (right panel, white rectangle). (c) Scenario of the unjamming accompanied with the grain motion via dilatancy. The seismogram in the plastically deformed zone reveals a group velocity of $V_G \sim 6$ m/s.

Nevertheless, by further increasing the driving amplitude, plastic deformation of the granular packing should become possible. To detect it, we examine the change in ultrasonic speckle patterns, i.e., B-mode images[28] (see the Supplemental Material) recorded before, during and after the passage of the nonlinear shear wave (Figs. 3a and 3b). As the characteristic length for rearrangement corresponds to the grain size $d$ (Fig. 3c), it falls in the spatial resolution of the ultrasound used here $\lambda_{US}/2 \sim 180$ μm ($\lambda_{US}$ is the wavelength in water). Fig. 3b (right panel) shows the case of a change in the speckle pattern for the large shear driving $F_{ac} \approx 2.7$ N ($\gamma_{ac} \sim 4.10^{-3}$) at a lower frequency of 100 Hz. Here the measured particle velocity reaches a very high value of $v_{ac} \sim 25$ mm/s and a particle displacement, more than $u_{ac}$

~ 40 μm, that becomes important compared to the grain size $d$ ~ 250 μm. This observation confirms the occurrence of an unjamming accompanied with a rearrangement of grain positions. The plastically fluidized zone appears close to the driving source within $x$ ~ 30 mm and along a depth $z$ ~ 7 mm as shown in the right panel of Fig. 3b. In the seismogram detected at the surface of the bead packing (Fig. 3c), two slopes are observed, respectively associated with the fluidized zone and unjammed zone without plastic rearrangement of grains. The shear velocity in the fluidized zone ($x$ < 30 mm) is reduced to 6 m/s!

**Discussion and modelling**

*Linear elasticity* –We use the mean-field approach as a guide to interpret our experimental data. According to the Biot theory for a fluid-filled granular porous medium, the saturating liquid increases the bulk modulus of the medium and couple compressional waves in the solid and liquid phases to form a fast and a slow mode[32]. This solid-liquid interaction is not expected to have a significant effect on the shear modulus of the solid skeleton but only to increase viscous dissipation between the solid and liquid phases[33]. A simple description of shear wave propagation in our granular suspension, can thus be given based on the effective medium theory[7,20,29] developed for dry random packing of frictional spheres (of radius $R$). Based on the affine approximation, the EMT has provided an adequate description of small-amplitude ultrasonic experiments in highly compressed granular solids[22], in situations where configuration-specific multiply scattered elastic waves do not probe any significant rearrangement of the contact force network. Such reversible sound-mater interaction[18,22] is also consistent with previous works where the linear response was observed at frictional interfaces due to pinned asperities[34]. For isotropic confining pressure $P$ (or load $W$), the bulk and shear moduli can then be related to the normal and tangential contact stiffness $k_n$ and $k_t$ as[35], $K \sim \phi Z k_n$ and $G \sim \phi Z (k_n + 3k_t/2)$ with $Z$ the average coordination number and $\phi$ the packing density of spheres (Fig. 4a). For the Hertz-Mindlin contact[8,35], the contact stiffnesses $k_n$ and $k_t$ (~ $k_n$) at small oscillation amplitude are related to the contact area (of diameter $a$,) and thus to the static compression $u_0$ ($\approx a^2/2R$) under a normal load $w$, $k_n \sim a \sim w^{1/3}$ (Fig. 4b); accordingly, on the macroscopic scale, $K$ and $G$ scale with pressure as ~ $P^{1/3}$. Compressional and shear wave velocities, given by $V_P = [(K + 4/3G)/\rho]^{1/2}$ and $V_S = (G/\rho)^{1/2}$, are thus expected to scale with $P^{1/6}$.

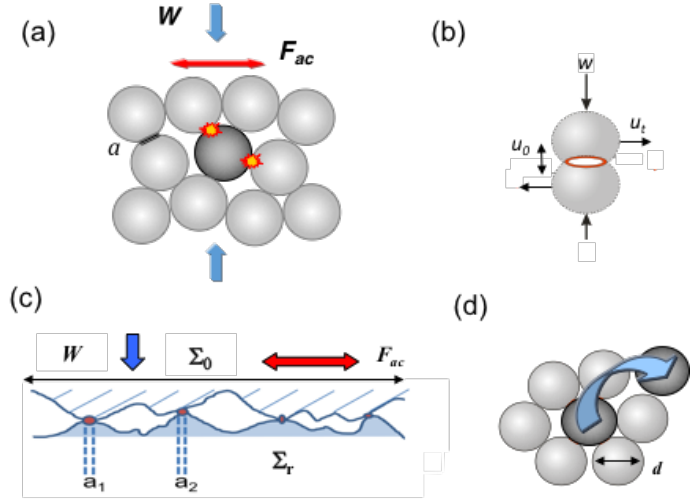

Fig. 4 (a) Confined elastic spheres packing (contact network) under shear. (b) The Hertz-Mindlin contact model. (c) Multi-contacts (asperities) formed between rough solid surfaces. (d) Scenario of the unjamming accompanied with the grain rearrangement.

However, for the shear acoustic measurements in weakly confined granular packings considered here, the affine approximation may break down even at relatively small amplitude of vibration due to induced slipping between grains or/and rearrangement of grain positions[7,8]. Consequently, the shear modulus can be overestimated by the EMT which does not allow the grains to relax via affine motion[3,8]. Numerical simulations show that the scaling of the linear shear modulus versus confining pressure shall be rewritten for both frictionless and frictional packings as[3,4,9],

$$G \sim K\,\Delta Z \sim k\,Z\,\Delta Z \qquad (1)$$

where $k$ is a linear combination of $k_n$ and $k_t$ and $\Delta Z = Z - Z_{iso}$ is the excess contact number (measuring the distance to isostaticity), which is related to the excess packing density by[3,9] $\Delta\phi \sim (\Delta Z)^2$ with $\Delta\phi = \phi - \phi_{iso}$. Note that the non-affine relaxation barely modifies the bulk modulus, i.e, the scaling remains as $K \sim P^{1/3}$ for the Hertz-Mindlin interaction. For a 3D packing of frictionless spheres ($Z_{iso} = 6$ and $\phi_{iso} = 0.64$), as both grain overlap (compression) $u_0$ and consequently $\Delta\phi$ scale as $\sim P^{2/3}$, one has $\Delta Z \sim P^{1/3}$, hence leading to $G \sim P^{2/3}$ and $V_S \sim P^{1/3}$. For frictional spheres, the packing does not tend to the isostatic value ($Z_{iso} = 4$) at unjamming under vanishing $P = 0$, but to a critical contact number $Z_c > Z_{iso}$[3,10]. The precise value of $Z_c$ depends on the friction between grains and the preparation history. The fact that $Z_c$ remains larger than $Z_{iso}$ explains why the shear modulus $G$ does not vanish at $P = 0$ (Eq. 1) and why it is possible to have a linear response of guided surface waves propagating along the free surface at small dynamic strain $\gamma_{ac} < 10^{-5}$ in our granular suspensions (Fig. 2b) and also in dry granular packings[11]. Indeed, even in the absence of an external load, compacted granular materials always have internal stresses which build up from friction between the grains (i.e., interlock) and from constraints imposed by the material boundary[36, 37].

*Nonlinear elasticity in frictionless packings* – We now address the nonlinear response of granular packings under finite shear beyond linear elasticity, either by steady[16] and oscillatory shear[15,17] or by nonlinear acoustics in this study. Our measurements show that the shear wave velocity, and accordingly modulus, softening in our frictional granular packing exhibits three different regimes, depending on the dynamic strain $\gamma_{ac}$. In regime (i), at small amplitude $\gamma_{ac}$ ($\leq 10^{-5}$), the wave velocity is constant; regime (ii) corresponds to higher driving amplitudes where the velocity decreases continuously without visible grain motion (Fig. 2b), down to the final regime (iii) accompanied by the plastic rearrangement of grains (Fig. 3c). To analyse these data, we first consider the extension of the shear modulus scaling law (eq. 1) beyond linear elasticity postulated by Otsuki and Hayakawa for a *frictionless* soft sphere packing, $G(\Delta Z, \gamma_0) \sim K \, \Delta Z \, \mathcal{G}[\gamma_0/(\Delta Z)^2]$ (we express $G$ as a function of $\Delta Z$ instead of $\Delta\phi$)[15]. Here $\mathcal{G}(x)$ is a scaling function with the asymptotic behaviour: $\mathcal{G}(x) \to$ constant when $x \to 0$ and $\mathcal{G}(x) \to x^{-1/2}$ when $x \to \infty$. The former recovers the linear response (Eq. 1) whereas the latter allows accounting for the shear modulus softening at large shear $\gamma_0$,

$$G(Z, \gamma_0) \sim K \, (\Delta Z)^2 \, \gamma_0^{-1/2} \qquad (2)$$

This scaling of $G$ with the amplitude $\gamma_0$ of the oscillatory shear can be explained based on an elasto-plastic model, which consists of an infinite number of connections in series with an elastic element of equal shear modulus $G_0$ and a slip element characterized by the drop stress $s$ (avalanche process). The stress of an individual element $\tilde{S}(s,t)$ (= $G_0 \, \gamma(t)$) is a linear function of the imposed strain $\gamma(t) = \gamma_0 \, (1 - \cos\omega t)$, but it drops to zero when exceeding the maximum value $s$ due to the breaking of the contact or bond (Fig. 4a). The shear modulus of the individual element is then calculated by $\tilde{G}(\gamma_0, s) = (-\omega/\pi) \int_0^{2\pi/\omega} \tilde{S}(s,t) \cos\omega t \, dt / \gamma$ and the global shear modulus is given by $G(\gamma_0) = \int_0^\infty ds \, \tilde{G}(\gamma_0, s) \rho(s)$. Here $\rho(s) \sim s^{-3/2} \exp(-s/s_c)$ is the probability density of the stress drop (with $s_c$ a characteristic stress), larger than the lower cutoff stress drop $s_0$ caused by the rearrangement of one grain. For $s_0/G_0 \ll \gamma_0 \ll s_c/G_0$, the shear modulus scales finally with $\gamma_0$ as $G \sim G_0^{1/2} \gamma_0^{-1/2}$ and is independent of $\omega$. On the other hand, Eq. 2 reveals a different power law for the scaling of the excess contact number $\Delta Z$ (or packing density $\Delta\phi$) than that of Eq. 1, which probably pertains to the rearrangement of grains via shear dilatancy caused by large shear[16]. However, the shear modulus softening observed in our experiments (Fig. 2b) is not necessarily associated with such plastic rearrangement.

*Nonlinear elasticity in frictional packings* –To specify the nonlinear elasticity of $G(\gamma_0)$ in realistic granular media, we propose a heuristic model where we replace the above elasto-plastic element by the Hertz-Mindlin frictional contact (Fig. 4b). Such contact is relevant not only between grains in granular media (Fig. 4a) but also in tribology and solid friction between asperities (Fig. 4c). Two distinct kinds of nonlinearity come into play at the contact area between two elastic spheres at large-amplitude vibration. In the normal direction, the Hertz contact law provides a relation between the oscillating force $f_n$ and the displacement $u_n$, $f_n \approx k_n u_n (1 + \beta u_n + O((u_n/u_0)^2))$ when $u_n \ll u_0$ (with the contact kept compressed). $\beta = 1/4u_0$ is the quadratic nonlinearity determined by $u_0 \sim w^{2/3}$. It is presumably responsible to the harmonics generation[23] (Fig. 2c) but affects little the normal stiffness $k_n^{NL}$ ($= f_n/u_n$) $\approx k_n (1 + O(u_n/u_0)^2)$. The other nonlinearity stems from the tangential friction where the Mindlin theory predicts both a nonlinear elasticity and dissipation from the hysteresis loop of force-displacement[22,38]. This hysteretic nonlinearity causes a weakening in the tangential stiffness $k_t^{NL}$ ($= f_t/u_t$) $\approx k_t (1 - f_t^*/6\mu w + O(f_t^*/\mu w)^2)$, proportional to the amplitude of the tangential force $f_t^*$ to lowest order ($\mu$ is the interparticle friction coefficient). The shear stiffness weakening is dominant[22] compared to that of the normal stiffness; it predicts a shear velocity softening for a moderately high shear $f_t^* < \mu w$ before reaching the yield of sliding,

$$\Delta V_S/V_S \sim (1/2)\Delta k_t/k_t \sim f_t^*/(6\mu w) \qquad (3)$$

which is about 10% for $f_t^* \sim F_{ac} \sim 1$ N, $w \sim W$ and $\mu \sim 0.2$. Unlike the mechanism of softening invoked in Eq. 2, the present slip-induced softening can occur with grains kept in contact during oscillation ($a > 0$ and $u_0 > 0$). This scenario may partly explain our observation of the dynamic softening up to $\Delta V_S/V_S \sim 40\%$ (Fig. 2b) without visible rearrangement of particles.

However, in a weakly granular packing compressed by gravity, the contact network is very inhomogeneous and the distribution of the (normal) contact force $w$ is exponential[8], $\rho(w) \sim exp(-w/w_c)$ with $w_c$ a characteristic force. This would give rise to the same distribution for the yield force $f_s = \mu w$ at the grain contacts, similar to the above drop stress distribution for the slip elements (Eq. 2). Under finite oscillatory shear, the hertzian contacts (or bonds) of smaller diameters $a \sim w^{1/3}$ will first break down via slipping which leads to the softening of interfacial shear stiffness but keeps the contacts overlapped. On the macroscopic level, the shear modulus softening or the unjamming sets on without rearrangement of the grain positions and consequently both the coordination number $Z$ and the packing density $\phi$ remains almost unchanged (Figs. 2b and 4a). Apparently, this softening process in a 3D sphere packing is comparable to what happens on a multi-contact interface under oscillatory shear[34] (Fig. 4c) where the distribution of the overlap $\rho(\delta)$ or compression $u_0$ (and the diameter $a$) between two asperities is also exponential.

To further investigate this similarity, we follow Bureau et al[34] to extend the Mindlin model for a single contact to the case of multiple micro-contacts in which we replace the asperities formed between 2D rough surfaces (Fig. 4c) by the bead contacts in a 3D granular packing. When the interface or the granular medium is loaded, the macroscopic shear force $F(t)$, or stress $\tau(t)$, is the sum of the contribution from all contacts $\rho(\delta)$. With a macroscopic shear force (or stress) of the form $F(t) = F_{dc} + F_{ac} \cos\omega t$, the shear displacement $U_t(t)$ (or strain $\gamma(t)$) is analytically derived using the distribution $\rho(\delta)$ over one cycle and the elastic response $U_{ac}$ (or $\gamma_{ac}$) is then obtained by $U_{ac} = (\omega/\pi) \int_0^{2\pi/\omega} U_t(t) \cos \omega t \, dt \approx 2\mu\lambda[(F_{ac}/2\mu W) + (F_{ac}/2\mu W)^2 + (5/4) (F_{ac}/2\mu W)^3 +...]$ with $\lambda$ a characteristic elastic length. The softening of the apparent shear stiffness by $k_t = F_{ac}/U_{ac}$ or shear modulus $G = \tau_{ac}/\gamma_{ac}$ may thus be written as function of the dynamic amplitude,

$$G/G_0 \sim k/k_{t0} = 1/[1 + F_{ac}/2\mu W + (5/4) (F_{ac}/2\mu W)^2] \quad (4)$$

where $G_0$ ($k_{t0}$) is the linear shear modulus (stiffness) at small $F_{ac}$. As expected for solid friction, the deduced shear modulus and stiffness are independent of $\omega$. The prediction from Eq. 4 shows a reasonably good agreement with the measured data, highlighting the crucial role of contact slipping in the shear modulus softening and the onset of unjamming[22,34,38]. The fitted friction coefficient $\mu \sim 0.12$ is a bit low but comparable to other measurements ($\sim 0.25$)[38].

*Plastic rearrangement* –Let us finally examine the shear velocity softening in the unjammed state accompanied with the rearrangement of the grain positions (Fig. 3b) produced at the strongest shear driving $\gamma_{ac} \sim 4.10^{-3}$. This nonlinear response is obviously associated with the change of the packing density $\Delta\phi$ and accordingly of the coordination number $\Delta Z \sim (\Delta\phi)^{1/2}$. Fig. 3c shows that the shear wave velocity is softened from an averaged value of 17 m/s at 100 Hz to a smallest value of 6 m/s at the fluidized zone close to the shear driving (< 20 mm). Such a huge velocity softening $\Delta V_G/V_G \sim \Delta V_S/V_S \sim 65\%$ corresponds to a shear modulus weakening ($G = \rho V_S^2$) of $\Delta G/G \sim 88\%$! We believe that the grain rearrangement via shear dilatancy (Fig. 4d) should also exist in the moderate nonlinear regime (Fig. 2b) where the rearrangement of grains may be too small ($< d/5 \sim \lambda_{US}/10$) to be detected due to the resolution of ultrasound imaging. These experiments evidence that the nonlinear shear response at finite acoustic strain is substantially plastic in the vicinity of unjamming transition where nonlinear elasticity cannot be decoupled from plasticity –a picture proposed by the simulation for athermal amorphous solids[39]. Further investigation on the shear modulus softening or unjamming is needed to quantify the interplay between vibration-induced contact slipping (Eq. 4) and contact breaking $\Delta Z < 0$ or shear dilatance $\Delta \phi < 0$ accompanied with the rearrangement (Eq. 2), and merge the two mechanisms into a unique model.

**Conclusion**

In summary, we have investigated shear wave propagation in weakly jammed dense granular suspensions. We monitored the unjamming transition by measuring the softening of wave velocity and shear modulus with increasing amplitude of the oscillatory shear. Towards unjamming transition, two successive process were found: contact slipping without any change of the packing density (and of the contact number) and plastic rearrangement of grains via shear dilatancy. Our measurements are consistent with the non-affine models in frictionless packings and agree particularly well with the extension of the Mindlin friction model on nonlinear elasticity, which evidences the important effects of contact slipping without contact loss. Such scenario of unjamming by acoustic fluidization/lubrication[23, 38, 40] should be helpful to better understand how transient seismic waves trigger avalanches and earthquakes[24] in sheared granular media.

**References**

1. Liu A.J. & Nagel S.R. Jamming is not just cool any more. *Nature* **396**, 21 (1998)
2. Cates M.E., Wittmer J.P., Bouchaud J.-P. & Claudin P. Jamming, force chains, and fragile matter. *Phys. Rev. Lett*. **81**, 1841 (1998)
3. O'Hern C.S., Silbert L.E., Liu A.J. & Nagel S.R. Jamming at zero temperature and zero applied stress: The epitome of disorder. *Phys; Rev. E* **68**, 011306 (2003)
4. van Hecke M. Jamming of soft particles: geometry, mechanics, scaling and isostaticity. *J. Phys.: Condens. Matter* **80**, 033101 (2010)
5. Bi D., Zhang J., Chakraborty B., Behringer R.P. Jamming by shear. *Nature* **480**, 355 (2011)

6. Brown & Jaeger H.M. Shear thickening in concentrated suspensions: phenomenology, mechanisms, and relations to jamming. *Rep. Prog. Phys*. 77, 046602 (2014)

7. Lerner, E., DeGuili, E., Düring G., and M. Wyart, M. Breakdown of continuum elasticity in amorphous solid. Soft Matter 10, 5085 (2014)

8. Makse H.A., Gland N., Johnson D.L., Schwartz L. Granular packings: nonlinear elasticity, sound propagation, and collective relaxation dynamics. *Phys. Rev. E* **70**, 061302 (2004)

9. Wyart M., Silbert L.E., Nagel S.R. & Witten T.A. Effects of compression on the vibrational modes of marginally jammed solids. *Phys. Rev. E* **72**, 051306 (2005)

10. Mizuno H., Mossa S. & Barrat J.-L. Acoustic excitations and elastic heterogeneities in disordered solids. *Proc. Natl. Acad. Sci. (USA)* **111**, 11949 (2014)

11. Bonneau L., Andreotti B. & Clément E. Evidence of Rayleigh-Hertz surface waves and shear stiffness anomaly in granular media. *Phys. Rev. Lett*. **101**, 118001 (2008)

12. Schreck C.F., Bertrand T., O'Hern C.S. & Schattuck M.D. Repulsive contact interactions make jammed particulate systems inherently nonharmonic. *Phys. Rev. Lett*. **107**, 078301 (2011)

13. Goodrich C., Liu A. & Nagel S. Contact nonlinearities and linear response in jammed particulate packings. *Phys. Rev. E* **90**, 022201 (2014)

14. Andreotti, B., Forterre, Y. and Pouliquen, O. *Granular Media: between fluid and solid* (Cambridge University Press, 2013)

15. Otsuki M. and Hayakawa H. Avalanche contribution to shear modulus of granular materials. *Phys. Rev. E* **90**, 042202 (2014)

16. Boschan J., Vagberg, D., Somfai, E., Tighe, B. Beyond linear elasticity: jamming solids at finite shear strain and rate. arXiv: 1601.00068v1 (2016)

17. Paredes J., Michels, M., and Bonn D. Rheology across the zero-temperature jamming. *Phys. Rev. Lett*. **111**, 015701 (2013)

18. Jia X., Caroli C. & Velicky B. Ultrasound Propagation in Externally Stressed Granular Media. *Phys. Rev. Lett.* **82**, 1863 (1999)

19. Goddard J.D. Nonlinear Elasticity and Pressure-Dependent Wave Speeds in Granular Media. *Proc. R. Soc. Lond. A* **430**, 105 (1990)

20. Khidas Y. & Jia X. Probing the shear-band formation in granular media with sound waves. *Phys. Rev. E* **85**, 051302 (2012)

21. Knuth M.W., Tobin H.J. & Marone C. Evolution of ultrasonic velocity and dynamic elastic moduli with shear strain in granular layers. *Granular Matter* **15**, 499 (2013)

22. Johnson P. and Jia X. Nonlinear dynamics, granular media and dynamic earthquake triggering. *Nature* **437**, 871 (2005)

23. Jia X., Brunet T. & Laurent J. Elastic weakening of a dense granular pack by acoustic fluidization: Slipping, compaction, and aging. *Phys. Rev. E* **84**, 020301I (2011)

24. Wildenberg S., van Hecke M. & Jia X. Evolution of granular packings by nonlinear acoustic waves. *Europhys. Lett*. **101**, 14004 (2013)

25. Gomez L.R., Turner A.M., van Hecke M. & Vitelli V. Shocks near jamming. *Phys. Rev. Lett* **108**, 058001 (2012); Ulrich S., Upadhyaya N., van Opheuseden B. & Vitelli V. Shear shocks in fragile networks. *Proc. Natl. Acad. Sci. (USA)* **110**, 20929 (2013) 11.

26. Nesterenko V.F. *Dynamics of Heterogeneous Materials* (Springer, NY, 2001)

27. Jacob X., Aleshin A., Tournat V., Leclaire P., Lauriks W., and Gusev V.E. Acoustic probing of the jamming transition in an unconsolidated granular medium. *Phys. Rev. Lett*. **100**, 158003 (2008)

28. Catheline S., Gennisson J.-L., Tanter M. & Fink M. Observation of shock transverse waves in elastic media. *Phys. Rev. Lett*. **91**, 164301 (2003)

29. Saint-Michel B., Bodiguel H., Meeker S., and Manneville S. Simultaneous concentration and velocity maps in particle suspensions under shear from rheo-ultrasonic imaging. *Phys. Rew. Appl*. 8, 014023 (2017)

30. - Wildenberg S, Jia X., Léopoldès J., and Tourin A. Ultrasonic tracking of a sinking ball in a vibrated dense granular suspension (*submitted*)

31. Metcalfe, G., Tennakoon, S., Kondic, L., Schaeffer, D., and Behringer, R. Granular friction, Coulomb failure, and the fluid-solid transition for horizontally shaken granular materials. *Phys. Rev*. **65**, 031302 (2002)

32. Bourbié T., Coussy O., and Zinsner B. *Acoustics of Pororus Media* (Edition TECHNIP, 1987)

33. Brunet T., Jia X. & Mills P. Mechanisms for acoustic absorption in dry and weakly wet granular media. *Phys. Rev. Lett*. **101**, 138001 (2008)

34. Bureau L., Caroli C., and Baumberger T. Elasticity and onset of frictional dissipation at a non-sliding multi-contact interface. *Proc. R. Soc. Lond*. *A* **459**, 2787 (2003)

35. Winkler K.W. Contact stiffness in granular porous materials: comparison between theory and experiment. *Geophys. Res. Lett*. **10**, 1073 (1983)

36. Khidas Y. and Jia X. Anisotropic nonlinear elasticity in a spherical–bead pack: influence of the babric anisotropy. Phys. Rev. E 81, 021303 (2010)

37. Del Gado E. Constructing a theory for amorphous solids. https://physics.aps.org/articles/pdf/10.1103/Physics.11.88 (2018)

38. Léopoldès L., Conrad G. & Jia X. Onset of sliding in amorphous films triggered by high-frequency oscillatory shear. *Phys. Rev. Lett*. **110**, 248301(2013)

39. Hentschel H., Karmakar S., Lerner E., and Procaccia I. Do athermal amorphous solids exist? *Phys. Rev. E* 83, 061101 (2011)

40. Melosh, H.J. Dynamical weakening of faults by acoustic fluidization. *Nature* **379**, 601 (1996)

**Acknowledgment**

We thank C. Caroli and M. Tanter for helpful discussions. J.B acknowleges the support from Instituto de Física, Facultad de Ciencias, Universidad de la República, Montevideo, Uruguay. This work was supported by French LABEX WIFI under references ANR-10-LABX-24 and ANR-10- IDEX-0001-02 PSL.

## Supplemental material

*Ultrafast Ultrasound Imaging*

To investigate the low frequency shear wave propagation and its effects on the granular packing, ultrafast ultrasound imaging was used. Through ultrasound it is possible to: i) image a cross sectional plane of the granular packing (as it is usually done in ultrasonography), and ii) measure the axial particle velocity field associated to shear wave propagation within the sample. To this end, the sample is insonified with ultrasonic waves emitted from an ultrasonic array. Each of the 192 elements of the array emits simultaneously a two-cycle short pulse centered at 4 MHz, thus generating a pulsed plane wave that propagates in the *xz*-plane in a direction perpendicular to the surface of the sample (Fig. S1a). This plane wave is scattered off the beads and the corresponding backscattered echoes are recorded by each element of the transducer array (Fig. S1b). The backscattered signal comes from the superposition of the echoes coming from different scatterers within the medium. Therefore, a beamforming step is necessary to construct an ultrasonic image, i.e. to relate the arrival time of an ultrasound echo to a given position within the imaging plane. In this work a standard parallel beamforming algorithm was used. Each point (*x,z*) of the image is obtained by adding coherently all the contributions coming from it. To that goal a time delay of the form $T(x,z) = \sqrt{z^2 + (x - x_i)^2}/c$ is first applied to the backscattered signals, where $x_i$ corresponds to the position of the *i*th element of the ultrasonic array and *c* is the speed of sound (= 1500 m/s in water) (Fig. S1c). Then, the 192 time-delayed backscattered signals are summed. Finally, these two steps are repeated for all points within the imaging plane to generate a beam-formed image as the one shown in Fig. 1d.

This beam-formed image is a cross-sectional image of the sample. When applying a logarithmic compression to increase contrast, a steel plate immersed in the granular medium appears as a strong echo at $z \approx 40$ mm and $x \approx 12$ mm (Fig. S1e), which shows that ultrasonic backscattering is still dominated by single scattering. That is, the arrival time of the speckle signal corresponds to a specific location of the scatterers in space. The comparison between such images recorded before, during and after the shear wave passage was used in this work to study grain rearrangement under high shear driving amplitude (see Figs. 3a and 3b).

The axial particle velocity field $v_z$ (*x, z, t*) (= $v_{ac}$) associated to shear wave propagation at a given instant *t* after the source excitation can be obtained (at least for the first ~10 mm of depth) by correlating in time the speckle pattern observed from one image $S_t$ (*x, z*) to the next $S_{t+\Delta t}$ (*x, z*) (i.e., by speckle interferometry)[29], with $\Delta t \sim 1/f_{RF}$ and $f_{RF}$ = 16 kHz the frame rate.

In practice, the operation is performed after quadrature demodulation of the RF backscattered signals. The field of the axial particle velocity $v_z$ at a particular time corresponding to the $n^e$ frame is then inferred from[S1]:

$$v_z(x, z, n) = \frac{f_{RF}}{2} \frac{c}{\omega} atan\left(\frac{Q(n)I(n+1)+Q(n+1)I(n)}{I(n)I(n+1)+Q(n+1)Q(n)}\right) \quad (5)$$

with $\omega$ the central frequency of ultrasound, $I$ and $Q$ the in-phase and quadrature-phase components of the demodulated signal corresponding to pixel $(x,z)$ in the image. As an example, a snapshot of the axial particle velocity field is presented at $t = 9.8$ ms in Fig. S1f. The map of the particle velocity as a function of time at $x = 42.5$ mm shows the ability to follow shear wave propagation in depth (Fig. S1g). Finally, Fig. S1h and Fig. 1c (in the main text) map the pulsed particle velocity as a function of $x$ and time $t$ at the surface of the sample.

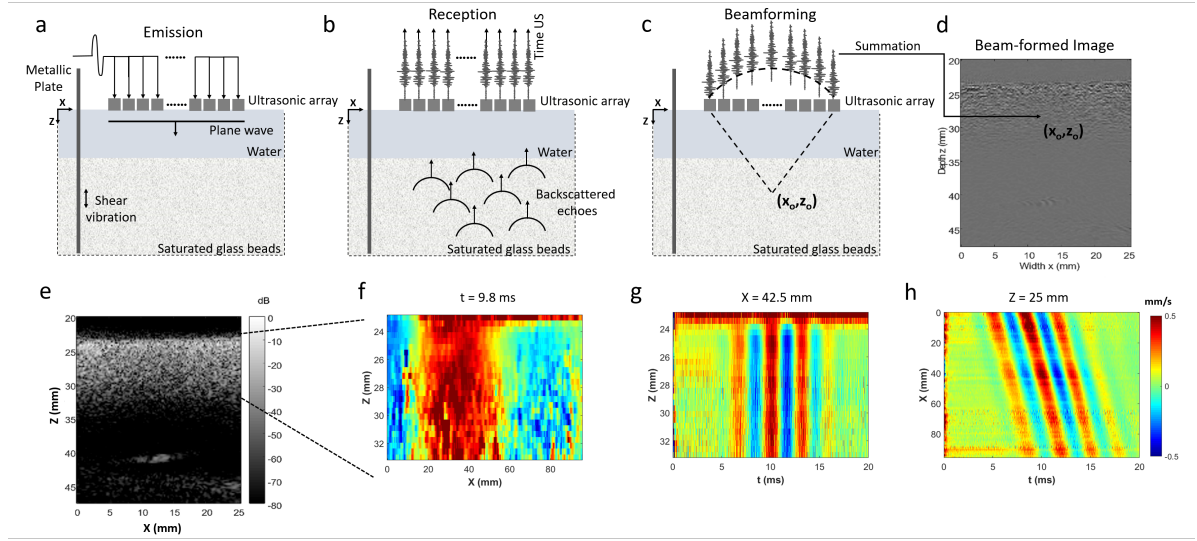

Figure S1 a) Ultrasound emission step: all elements of the array emit simultaneously a short pulse centered at 4 MHz, thus generating a pulsed plane wave b) Ultrasound reception step: the backscattered echoes coming from different locations within the medium are recorded by each element of the transducer array c) Beam-forming step: to relate the arrival time of an ultrasound echo to a given position within the imaging plane, each point $(x_o, z_o)$ of the image is obtained by adding coherently the backscattered signals originating from it. e) Logarithmic compression of the beam-formed image to increase image contrast. A steel plate immersed in the granular medium appears as a strong echo at $z \approx 40$ mm and $x \approx 12$ mm. f) One snapshot of the axial particle velocity field associated to shear wave propagation at $t = 9.8$ ms after the generation of the shear wave. g) Particle velocity as a function of depth and time at $x = 42.5$ mm. h) Axial particle velocity field as a function of time along $x$ at the surface of the sample.

*Guided Acoustic Modes*

Bonneau et al[11] and Jacob et al[27] have demonstrated that wave propagation along the free surface of a granular packing may be described by a superposition of localized acoustic modes. To define which guided acoustic modes are generated with our setup (Fig. 1a in the main text) we conducted a series of numerical finite element method simulations (FEM) with COMSOL Multiphysics, following a similar approach as proposed by Bergamo et al[S2]. We model the granular packing in the long wavelength limit ($\lambda_{LF} \gg d$) as a continuous elastic layer of thickness $h = 130\ mm$. A gravity-induced stiffness gradient was included in the FEM simulations given by the scaling law $V_{P,S} = \gamma_{P,S}.(\rho g z)^{\alpha_{P,S}}$, where $V_{P,S}$ is the compressional/shear wave velocity, $\gamma_{P,S}$ is a depth-independent coefficient, $\alpha_{P,S}$ is the power law exponent, $\rho$ is the bulk density of the medium, $g$ is the gravity acceleration and $z$ is the depth. For the simulations we used $\rho$ = 1700 kg/m$^3$, $\alpha_S$ = ¼ and $\alpha_L$ = 1/6 as in Refs. [8, 19]. Finally, $\gamma_S$ and $\gamma_L$ were chosen equal to 5.25 and 14.8 respectively as reported in Ref. [27] as best fit parameters for a dry granular packing. The boundary conditions were set as free at the surface and clamped at the bottom. Experimentally, waves are generated using a rough metallic plate acting all across the layer sample (Fig. 1a in the main text). Therefore, in the FEM simulations a prescribed displacement was assigned to the plane $x$ = 0 with the same temporal dependence used in the experiments: a four-cycle tone-burst centered at a frequency between 100 and 500 Hz.

Figure S2 shows three snapshots of the vertical component of the simulated particle velocity field in the *xz*-plane for a 300 Hz excitation. Two different modes of propagation are generated with our setup: a first mode localized near the free surface and a second mode that propagates within the bulk of the sample (Mode 1 and 2 respectively in Fig. S2). Due to the size of the imaging plane used in the experiments, the particle velocity is presumably dominated by Mode 1 as can be verified in Fig. S3 that shows experimental and simulated particle velocity fields. Experiment and simulation are found in good agreement. Furthermore, the presence of a single mode of propagation may be observed in the corresponding double Fourier transformation (Figs. S3c and S3d).

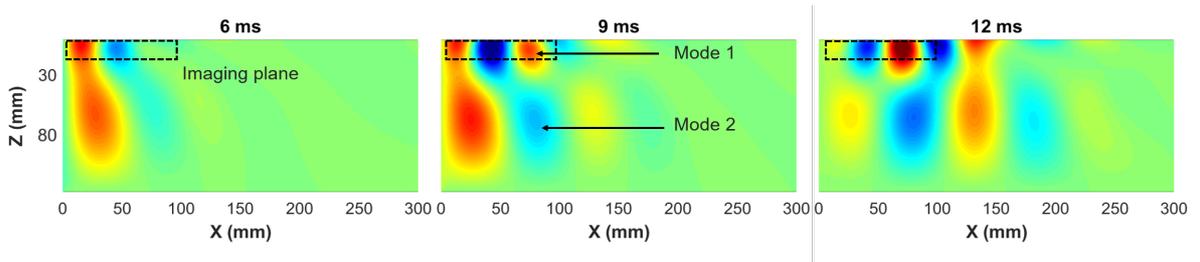

Figure S2 Three snapshots of the vertical component of the simulated particle velocity field during shear wave propagation for a 300 Hz excitation. The black dashed rectangle corresponds to the imaging plane used in the experiments presented in Fig.S1 f-h.

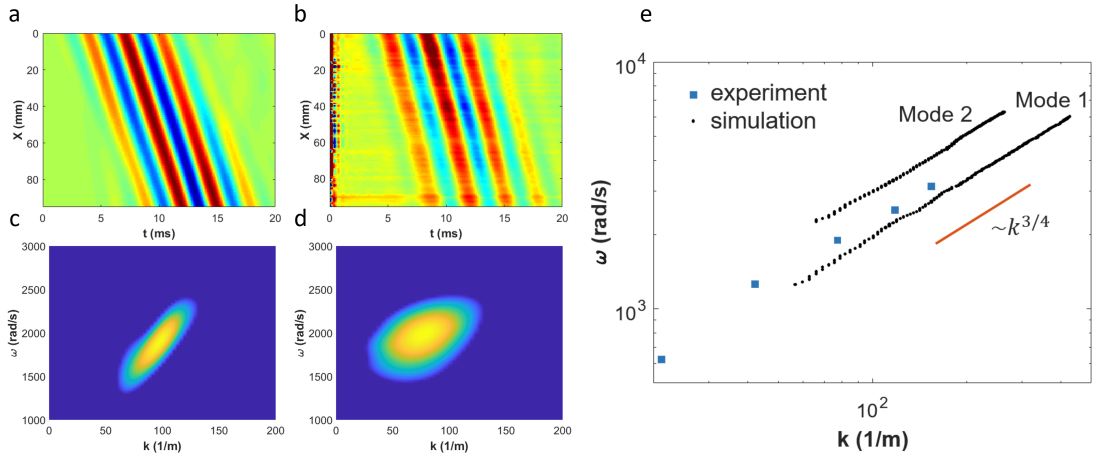

Figure S3 Particle velocity field for a four cycle tone burst excitation centered at 300 Hz in a) simulation and b) experiment along with its double Fourier Transform in c) and d) respectively. e) ω-$k$ relation extracted from experiment (blue squares) and simulation (black dots). The relation $\omega \propto k^{3/4}$ is plotted in red full line for reference.

The dispersion relation (depicted on a log-scale in Fig. S3e for both simulation and experiment) helps to understand the nature of this mode. With simulation it was also possible to calculate the *ω-k* relation within the bulk of the sample at *z* = 80 mm (mode 2 in Fig. S2). From this figure it is possible to establish that *ω* reasonably follows a relation proportional to $k^{3/4}$ (an exponent of 0.79 was found by fitting the experimental data in Fig. S3). Therefore, the mode detected by our setup corresponds to the lowest surface mode described by Jacob et al. This type of localized surface mode is controlled by the shear wave speed profile and is reminiscent of a Rayleigh wave in an elastic homogeneous layer (i.e. without velocity gradient). Consequently, its propagation velocity is directly linked to the shear modulus of the sample.

**References**

S1. Loupas T, Powers J.T., Gill R.W. An axial velocity estimator for ultrasound blood flow imaging, based on a full evaluation of the Doppler equation by means of a two-dimensional autocorrelation approach. *IEEE Trans Ultrason Ferroelectr. Freq. Control*. 42, 672 (1995)

S2. Bergamo P. et al. Physical modelling of a surface-wave survey over a laterally varying granular medium with property contrasts and velocity gradients, *Geophys. J. Int*. **197**, 233 (2014)